\documentstyle[12pt]{article}
\begin{document}

\textheight 25cm

\vspace*{3cm}
\begin{center}
STRUCTURE AND DECAYS OF PARTICLES CONTAINING HEAVY QUARKS (THEORY)\\
\vspace*{1cm}
Kacper Zalewski\footnote{Also at the Institute of Nuclear Physics, Cracow. This
work was partly supported by the\\ KBN grant 2P 302 076 07}\\
Institute of Physics, Jagellonian University, ul. Reymonta 4\\
Poland 30 059 Krak\'ow\\
\end{center}

\vspace*{4cm}

Some recent results and problems in the theory of particles containing heavy
quarks are reviewed.

\newpage

\section{Introduction}

The subject of this review is the status of the theory of particles containing
heavy quarks, except for production processes, which will be discussed by
P.Nason$^{1)}$ in the next talk. Let us note first that there is much
activity in this field of research. In the high energy physics data base {\it
hep-ph}, in 1994 I have found more than four hundred papers on this subject.
Out of that about 15\% were concerned with production processes, but what is
left is more than one paper per day including sundays. A linear extrapolation
of the number of papers submitted to this data base in the first five months of
this year leads to an expectation of over five hundred papers in 1995; thus the
activity keeps increasing. This implies that in my presentation I shall have to
be very selective.

I shall discuss five problems.
\begin{itemize}

\item Determination of the fundamental parameters of the standard model. This
is perhaps the most often invoked motivation for the study of heavy quark
physics. Ten of these parameters have names of heavy quarks in their symbols
($m_c,\; m_b,\; m_t,\; V_{ub},\; V_{cd},\;V_{cs},\; V_{cb},\;
\\V_{td}, V_{ts} \mbox{ and } V_{tb}$). Moreover, some others can be determined
using heavy quark systems. An example is the determination of $\alpha_s$ from
the decays of quarkonia, which has been discussed at this conference by M.
Schmelling$^{2)}$.

\item The heavy quark expansions. Such expansions, there is more than one, have
been a break through in heavy quark physics. They were the subject of a
separate talk at the previous Physics in Collision Conference$^{3)}$.

\item CP nonconservation and mixing. Here the predictions of the standard model
are particularly stringent and "new physics" has perhaps a chance of becoming
visible.

\item Rare decays. Here the rapidly improving accuracy of the experimental data
and of the theoretical analyses has already provided a striking confirmation of
the standard model, but further progress is possible and will lead to important
tests of the model.

\item Nonleptonic decays. This is a difficult problem$^{4)}$, where progress
since 1987 has been slow and new ideas seem to be particularly needed.

\end{itemize}

\section{Quark masses}

The values of the quark masses are among the fundamental parameters of the
standard model and have been used for many years and in many calculations.
Nevertheless; their values are poorly known and controversial. In the Particle
Data Group Tables they were listed for the first time in 1992$^{5)}$ and
even then with the comment that they are given "to provoke discussion". The
values proposed for the heavy quark masses at that time were $m_c = (1.3
\mbox{---} 1.7)$ GeV, $m_b = (4.7 \mbox{---} 5.3)$ GeV and $m_t > 91$ GeV.
Since then the discovery of the $t$ quark has been a break through on the
experimental side$^{6)}$, but also on the theoretical side much
clarifying work has been done.

It has been realized, how important is the question "what do you mean by the
quark mass?". In order to apply the standard definition $m = \sqrt{E^2 -
\vec{p}^2}$ one would have to perform the measurements on a free quark. This,
however, according to quantum chromodynamics (QCD) is impossible. The next
choice is the use of the term $m\overline{\psi}\psi$, which occurs in the QCD
Lagrangian and contains explicitly the quark mass. This, however is the bare
mass, which is not measurable. Related to measurable quantities is the
self-energy, which is the sum of the bare mass and of corrections corresponding
to Feynman diagrams containing loops. The first correction corresponds to the
diagram, where the quark emits and reabsorbs a virtual gluon. It reads,
see e.g.$^{7)}$,

\begin{equation}
\Sigma^{(1)} = \frac{m \alpha_s(\mu)}{\pi}\left[ \frac{1}{\varepsilon} - \gamma
+ \log(4\pi) +\log\frac{\mu^2}{m^2} + \frac{4}{3} \right],
\end{equation}
where $\gamma = 0.5772\ldots$ is the Euler constant and the limit $\varepsilon
\rightarrow 0$ is understood. This formula clearly exhibits the two main
difficulties. The first term in the square bracket is infinite and the scale
factor $\mu$ is arbitrary. The remedy is to rewrite the self-energy in the form

\begin{equation}
\label{sig1lo}
m + \Sigma^{(1)} = (m + \delta m) + (\Sigma^{(1)} - \delta m)
\end{equation}
and to choose $\delta m$ so that $\Sigma^{(1)} - \delta m$ is finite. Since the
bare quark mass $m$ is unmeasurable, one can assume that the new quark mass

\begin{equation}
m_Q = m + \delta m
\end{equation}
is finite. This is, of course, the standard mass renormalization procedure.
Including more Feynman diagrams, one has to redefine $\delta m$ and $m_Q$, but
let us consider first the one gluon loop result (\ref{sig1lo}). Many choices of
$\delta m$ are possible. For instance, choosing $\delta m$ to cancel the
$\frac{1}{\varepsilon}$ term only one obtains the $MS$ (minimal subtraction)
mass, choosing $\delta m$ to cancel also the $-\gamma + \log(4\pi)$ term one
obtains the so called $\overline{MS}$ mass and choosing $\delta m$ to cancel
all the $\Sigma^{(1)}$ expression one obtains the pole mass. Using elementary
algebra one can correlate the various mass definitions. For instance, the
relation (in the one gluon loop approximation!) between the $\overline{MS}$
mass calculated for $\mu = \overline{m}_Q$, further called the running mass,
and the pole mass is

\begin{equation} \overline{m}_Q(\overline{m}_Q) = m^{Pole}_Q\left(1 - \frac{4
\alpha_s(\overline{m}_Q)}{3 \pi} \right). \end{equation} The differences
between the pole masses and the running masses are very significant. From the
simple formula given above, substituting typical values $\alpha_s(m_c) \approx
0.35$, $\alpha_s(m_b) \approx 0.20$ and $\alpha_s(m_t) \approx 0.10$, one finds
the differences $m^{Pole}_Q - \overline{m}_Q(\overline{m}_Q)$ approximately
equal to $0.17$ GeV for the $c$-quark, $0.34$ GeV for the $b$-quark and $7$ GeV
for the $t$-quark. A more careful calculation by Titard and Yndurain$^{8)}$
gives $0.26$ GeV for the $c$-quark and $0.51$ GeV for the $b$-quark. For the
$t$-quark typical values are ($8\mbox{---}9$)GeV.

The recent Particle Data Group Tables$^{9)}$ contain a discussion of
several definitions of quark masses and choose for the $c$-quark and for the
$b$-quark the masses $\overline{m}_Q(\overline{m}_Q)$ to put into the tables.
They propose $m_c = (1 \mbox{---} 1.6)$ GeV and $m_b = (4.1 \mbox{---} 4.5)$
GeV. The uncertainties remain very large, but the definition is now clear. An
obvious question is, how the newly measured mass of the $t$-quark$^{6)}$
$m_t \approx 174$ GeV should be interpreted? Even with the present
uncertainties, the almost $10$ GeV difference between the pole mass and the
running mass for the $t$-quark is not negligible. In principle, the
experimental procedure defines a new mass $m_t$. In practice, however,
everybody expects that this definition is close (to better than one GeV say) to
the pole mass or to the running mass. Analogies with resonance decays and the
fact that the $t$ and $\overline{t}$ at the Tevatron are usually produced in
the colour octet state, i.e. do not interact very much with each other, suggest
the pole mass, but arguments in favour of the running mass are also quoted. The
problem is not yet quite solved. Here we would only like to stress its
importance.

Taking into account higher orders of perturbation theory one finds an
interesting difficulty in connection with the definition of the pole mass. The
series used to define this mass is found to diverge$^{10),11),12)}$. This is
usually described in terms of renormalons, which have been discussed at this
conference by M. Karliner$^{13)}$. The renormalons can be eliminated in favour
of terms divergent like powers of the cutoff, or terms divergent like inverse
powers of the the lattice spacing, but the final result is always the same ---
the pole mass cannot be defined better than with an uncertainty of about $50$
MeV. A recent review of this problem has been given by Sachrajda$^{14)}$. There
are two ways out of this difficulty. One is to abandon the pole masses and to
express everything in terms of running masses. This is being advocated e.g. by
Bigi and collaborators$^{11)}$. Another way is to introduce nonperturbative
subtractions, which tame the divergence and enable a precise definition of a
"subtracted pole mass", as proposed by Martinelli and Sachrajda, see$^{14)}$
and references given there.

\section{Heavy quark expansions}

The heavy quark expansions use the fact that the masses of the heavy quarks are
much larger than the low energy mass scales, $\Lambda_{QCD},\;\Lambda_{chiral}$
and $\overline{\Lambda} = M_H - m_Q$, where $M_H$ denotes the mass of a hadron
containing one heavy quark $Q$. An extensive review has been given in Physics
Reports by Neubert$^{15)}$. Very recent reviews include the review by
Bigi$^{16)}$ for inclusive processes and by Mannel$^{17)}$ for exclusive
processes. At this conference applications have been presented by P.
Jarry$^{18)}$ and R. van Kooten$^{19)}$.

Already the zero order approach, where $m_Q^{-1} \rightarrow 0$, leads to many
beautiful results, like the prediction that

\begin{equation}
M_{B^*} - M_B \ll M_{B^{**}} - M_B,
\end{equation}
or the expression of the six form factors necessary to describe the
semileptonic decays $B \rightarrow D^{(*)}l\nu$ in terms of a single Isgur-Wise
function $\xi(\omega)$, where $\omega$ is the Lorentz factor of the
$D^{(*)}$-meson in the rest frame of the $B$-meson. Since this has been
reviewed many times, however, it may be more interesting to discuss the
difficulties of the heavy quark approach. Difficulties is here not an euphemism
for disagreement with experiment; we shall discuss problems, which required, or
still require hard work and/or new ideas. We shall discuss three problems
\begin{itemize}
\item Corrections from quantum field theory (QFT),
\item How to compare theory with experiment in spite of the occurrence of
arbitrary functions in the theoretical expressions,
\item Corrections from higher orders in the expansion in powers of $m_Q^{-1}$.
\end{itemize}

\subsection{QFT corrections}

A nice introduction to the methods of calculating the quantum field theory
(QFT) corrections can be found in Neubert's review$^{15)}$. The present
status of such calculations for weak decays has been very recently summarized
by Buras$^{20)}$. The calculations are often hard work ---  years of work
for teams of experts. They are nevertheless essential for a quantitative
comparison of the theory with experiment. In particular, they reduce or
eliminate the dependence of the results on the scale parameter $\mu$. Much
remains to be done. For example, the difference between the pole mass and the
running mass of the $t$-quark is known with an uncertainty much larger than
that caused by renormalons, because the necessary QFT corrections have not yet
been calculated.

\subsection{Arbitrary functions}

In order to illustrate how the arbitrary functions enter the predictions, let
us consider the semileptonic decay $B \rightarrow D^{(*)}l\nu$. In the heavy
quark picture this process can be decomposed into two steps. First the $b$
quark emits the lepton pair and goes over into the $c$-quark. This happens at a
high energy scale, where the light components of the mesons are ineffective.
The matrix element for this part of the process is calculated as if the heavy
quarks were free. Then the light component of the $B$-meson with velocity
$\vec{v}$ must reorganize itself into the light component of the $D^{(*)}$
meson recoiling with velocity $\vec{v}'$. The probability amplitude for that to
happen is (almost) unknown and is the Isgur-Wise function $\xi(\omega)$, where
$\omega = v^\mu{v_\mu}'$. Thus, the decay amplitude for the decay is
proportional to

\begin{equation}
A = V_{cb} \overline{u}_{\vec{v}'}\gamma_\mu(1-\gamma^5)u_{\vec{v}}\xi(\omega).
\end{equation}

A typical problem is how to extract from the experimental data the
Cabibbo-Kobayashi-Masakawa ($CKM$) matrix element $V_{cb}$ in spite of the
unknown factor $\xi(\omega)$. Two approaches to this problem have been popular.
In the exclusive approach, reviewed recently by Neubert$^{21)}$, one uses the
fact that in the case of no recoil $\vec{v} = \vec{v}'$, or equivalently
$\omega = 1$, the light components of the $B$-meson and of the $D^{(*)}$-meson
are almost the same. Thus $\xi(1) \approx 1$. The corrections are small and
calculable. In the inclusive approach reviewed recently by Bigi$^{16)}$, one
uses the fact that the sum of probabilities for all the possible rearrangements
of the light component must be equal one. Thus, in each case one identifies an
experimental quantity --- the semileptonic decay probability to $D^{(*)}$ at
zero recoil in the exclusive method, the total semileptonic decay width in the
inclusive method --- for which the theoretical prediction does not depend on
the arbitrary function. So the $CKM$ matrix element $V_{cb}$ can be extracted
in a model independent way.

An obvious question is, which method is better? Since both methods are applied
to the same experiments, a good measure of their quality is the error on
$V_{cb}$. Neubert$^{21)}$ quotes for the exclusive method the error $\pm 0.003
\pm 0.002$ and for the inclusive method $\pm 0.001 \pm 0.005$, where in each
case the first error is experimental and the second is theoretical.
Bigi$^{16)}$ finds for the corresponding errors $\pm 0.003 \pm 0.002$ and $\pm
0.002 \pm 0.002$. A comparison of these numbers shows that experimentally the
inclusive method is easier. There is a controversy, however, concerning the
theoretical uncertainties in the inclusive method. In practice, of course, one
should use both methods and average the results.

Let us note finally that interesting results can also be obtained by studying
the inclusive processes in the small recoil velocity limit$^{22),23)}$, which
combines the inclusiveness with the selection of a simplifying kinematical
configuration.

\subsection{Higher orders in $m_Q^{-1}$}

The limit $m_Q^{-1} \rightarrow 0$ gives many beautiful results, but sometimes
it is too crude. For instance, we may not be satisfied with the prediction
$M_{B^*} \approx M_B$. Assuming that the corrections are proportional to
$m_Q^{-1}$ one gets immediately

\begin{equation}
\frac{M_{B^*} - M_B}{M_{D^*} - M_D} \approx \frac{m_c}{m_b},
\end{equation}
which is very reasonable.

The problem is to find a systematic expansion of the quantities of interest
into inverse powers of the quark mass. For simplicity we shall use the language
of quantum mechanics, but most of the results presented here are available also
in full quantum field theory. One replaces the bispinors $\psi$ describing the
heavy quarks by new bispinors $\psi'$ defined by the relation

\begin{equation}
\psi = U\psi'
\end{equation}
This corresponds to the replacement of the $QCD$ Lagrangian ${\cal L}$ by a new
Lagrangian ${\cal L}$' so that

\begin{equation}
\overline{\psi}{\cal L}\psi \equiv \overline{\psi}'{\cal L}'\psi'.
\end{equation}
Any current can be rewritten in terms of the transformed bispinors e.g.

\begin{equation}
\overline{\psi}\Gamma^\mu\psi = \overline{\psi}'\gamma^0 U^{\dag} \gamma^0
\Gamma^\mu U \psi'.
\end{equation}
The problem would be solved, if one could choose the matrix $U$ so that the
bispinors $\psi'$ do not depend on the quark masses, while the operator
$\gamma^0 \ldots U$ is a power series in inverse powers of the heavy quark
mass. In practice the matrix $U$ is built in a series of steps.

First, a transformation is chosen so that the bispinors $\psi'$ satisfy the
relation

\begin{equation}
\psi' = (\gamma^\mu v_\mu) \psi',
\end{equation}
where $v$ is the velocity of the hadron containing the heavy quark. In the rest
frame of the hadron this simply means that only the first two components of the
bispinor can be different from zero. This step is easy and there are various
ways of performing it. Thus, Georgi$^{24)}$ uses the equations of motion to
eliminate the "small" components of the bispinor, K\"orner and Thompson$^{25)}$
use an analogue of the Foldy-Wouthuysen transformation and Mannel and
collaborators$^{26)}$ use the functional formalism.

The second step is to ensure a normalization of the bispinor $\psi'$, which
does not depend on mass, e.g. a normalization to unity. In the Foldy-Wouthuysen
approach this is automatic. In the others one has to introduce a
suitable normalizing factor, but this also is not difficult.

Finally, it is necessary to make the two independent components of $\psi'$
independent of the quark mass and here there is no simple method. In practice
one uses perturbation theory to express the original bispinors $\psi'$, which
are solutions of the equations of motion corresponding to the Lagrangian ${\cal
L}$' and inherit its dependence on mass, by solutions of the equations of
motion with an approximate Lagrangian ${\cal L}_0$', which does not introduce a
mass dependence into the solutions. Perturbative expressions, however, become
very complicated, when one goes to higher orders. Therefore, in practice it is
difficult to go beyond one or two nonvanishing corrections to the leading term.

Let us summarize the situation. The leading terms are well understood and
rather simple. For low order corrections and suitably chosen quantities it is
often possible to express the quantities of interest in terms of a few
constants with clear physical interpretations. A much discussed constant, which
occurs at order $m_Q^0$ in mass calculations, is

\begin{equation}
\overline{\Lambda} = M_H - m_Q
\end{equation}
Since the hadron mass $M_H$ is measurable, a discussion of this constant
reduces to the discussion of the quark mass, slightly complicate by the fact
that $\overline{\Lambda}$ should not depend on the quark mass. In the next
order of the mass calculations two more constants appear

\begin{eqnarray}
\mu^2_G & \equiv & 3 \lambda_2 \approx \frac{3}{4}(M_{B^*}^2 - M_B^2) \approx
0.37 GeV^2,\\
\mu_\pi^2 & \equiv & -\lambda_1 = <\vec{p}_Q^2>  \approx 0.37 \mbox{---} 0.75
GeV^2,
\end{eqnarray}
We have quoted here both of the two most popular notations. These constants
have simple physical interpretations. The first corresponds to the interaction
of the chromomagnetic moment of the heavy quark with the chromomagnetic field
produced by the light component. This interaction is responsible for the mass
splitting between the $B$ and the $B^*$, therefore it can be reliably
determined from experiment. Since for a spin one half fermion its
(chromo)magnetic moment is inversly proportional to its mass, the constant
appears at order $m_Q^{-1}$. The second constant is related to the kinetic
energy of the heavy quark in the hadron. In the rest frame of the hadron, the
momentum of the heavy quark is equal to the momentum of the light component and
does not depend (to zero order!) on the flavour of the heavy quark. The kinetic
energy has, however, an additional factor $m_Q^{-1}$, which introduces a
flavour dependence at this order. This constant is not directly measurable. A
lower limit for it

\begin{equation}
\mu_\pi^2 \geq \mu_G^2
\end{equation}
has been found in refs$^{27),28),23)}$. The upper limit given in formula (14)
is just a popular guess. The higher order calculations contain so much
arbitrariness that new ideas may be needed to make them useful.

\section{Mixing and CP nonconservation}

According to the standard model CP nonconservation and its effects in
particle--antiparticle mixing result only from the fact that the $CKM$ matrix
is not real. In terms of the parameter $\lambda \approx \frac{1}{5}$, the
orders of the real and imaginary parts of the elements of the $CKM$ matrix are
as follows

\begin{equation}
V \approx \left( \begin{array}{ccc}
                 1 & \lambda & (\lambda^3, i\lambda^3)\\
                 (-\lambda, i\lambda^5) & (1, i\lambda^5) & \lambda^2 \\
                 (\lambda^3, \lambda^3) & (\lambda^2, i\lambda^5) & 1
                 \end{array} \right).
\end{equation}
Here, as usual, the columns correspond to the quarks $d,\;s,\;b$ and the rows
to the quarks $u,\;c,\;t$. Thus, e.g. the entry in the upper right corner
indicates that the matrix element $V_{ub}$ has both the real and the imaginary
part of order $\lambda^3$. This element and $V_{td}$ are the only two, which
have comparable real and imaginary parts. The imaginary parts of the other
elements are either zero, or very small. It is also useful to keep in mind the
experimental uncertainties of the elements. These are summarized below

\begin{equation}
\delta V \approx \left( \begin{array}{ccc}
                 0.06 & 1.3 & 43\\
                 1.3 & 0.07 & 20 \\
                 58 & 20 & 0.035
                 \end{array} \right)\%.
\end{equation}
All these data are from the latest Particle Data Group Tables$^{9)}$. They
have been obtained using the unitarity of the $CKM$ matrix

\begin{equation}
V V^{\dag} = 1.
\end{equation}
It is only because of unitarity that $V_{tb}$ is the best known number in the
table. The unitarity of the $CKM$ matrix $V$ implies also that the scalar
product of any two rows, as well as the scalar product of any two columns,
vanishes; for $j \neq k$:

\begin{equation}
\sum_i V^*_{ji}V_{ki} =0; \hspace{1cm} \sum_i V^*_{ij}V_{ik} =0.
\end{equation}
The fact that the sum of three complex numbers vanishes means geometrically
that the corresponding vectors form a triangle in the complex plane. There are,
therefore, six unitarity triangles corresponding to the three pairs of rows and
the three pairs of columns. Four out of these triangles are almost degenerate,
with one side much shorter than the other two. The other two contain each
essentially the same information, thus it is enough to use one of them, e.g.
the triangle corresponding to the product of the first and the third column.
The complex number $V_{cd}^*V_{cb}$ is almost purely real and reasonably well
known. Therefore, it is usual to divide the lengths of all sides by the
absolute value of this number and thus to obtain a reduced triangle with
vertices at the points (0,0) and (1,0). The challenge is to obtain the two
coordinates of its third vertex.

Knowing the position of this vertex one can in principle, i.e. assuming some
values for a number of less fundamental parameters like decay constants, bag
constants, phase shifts defining final state interactions etc., predict all
the asymmetries and mixings in heavy quark systems. So many predictions can be
made that this looks like a very stringent test of the standard model and a
good place to look for new physics.

For the moment our knowledge about the position of the third vertex is rather
vague. Let us denote by $\alpha,\;\beta\;\mbox{ and }\gamma$ the angles of the
triangle at the vertices $(0,0)$, $(1,0)$ and at the third vertex respectively.
The knowledge of these angles is, of course, equivalent to the knowledge of the
position of the third vertex. A recent analysis of the available experimental
data performed by Pich and Prades$^{29)}$ has given the following limits for
these angles

\begin{eqnarray}
-0.23 & \leq & \sin(2 \alpha) \leq 1.0\\
 0.51 & \leq & \sin(2 \beta)  \leq 0.86\\
-0.80 & \leq & \sin(2 \gamma) \leq 0.95
\end{eqnarray}
These bounds are not very sure --- in the same paper one can find another set,
though less favoured by the authors --- but they show that nontrivial
statements about the angles i.e. about the position of the third vertex are
just beginning to be possible. The field is of great interest and potential
importance. Incidentally, it should be kept in mind that the limits on the
single angles are somewhat misleading, because the angles are correlated.
Besides the obvious correlation $\alpha + \beta + \gamma = \pi$, there are also
important correlations between pairs of angles.

It is remarkable that the lengths of the sides of the triangle, which is
another way of saying the position of the third vertex, can be obtained from
processes unrelated to CP nonconservation cf. e.g.$^{30)}$. The absolute value
$|V_{ub}|$ can be obtained from studies of the decays $B \rightarrow X_u l
\nu$.The absolute value $|V_{td}|$ can be obtained either from the mass
difference $\triangle M_s = M_{B_sH} - M_{B_sL}$, or from the rare decays $B
\rightarrow X_d\gamma$. Unfortunately, none of the corresponding experiments is
easy. There seems to be a kind of complementarity at work here. When a
measurement is comparatively easy, like looking for charmless semileptonic $B$
decays, its theoretical interpretation is unclear. When theory provides a nice
prediction like$^{31)}$

\begin{equation}
\frac{\triangle M_s}{\triangle M_d} = (1.3 \pm 0.2)\left|
\frac{V_{ts}}{V_{td}} \right|^2,
\end{equation}
the necessary measurements are difficult. In this particular case $\triangle
M_d = (0.471 \pm 0.026)\times 10^{12}\hbar sec^{-1}$ is sufficiently
well known$^{19)}$, but $\triangle M_s$ is proportional to the parameter $x_s$,
for
which only a lower limit $x_s > 9$ is known from experiment$^{31)}$, while
the expectation is $x_s \approx 20$, which may be beyond the range even of HERA
B.

\section{Rare decays}

The name rare decays is applied to decays involving neutral flavour-changing
neutral currents. The most popular process is $b \rightarrow s\gamma$, for
which experimental data is available$^{32),33)}$. Such processes are forbidden
at the tree level and they are ascribed to pingwin diagrams, where the
$b$-quark dissociates into a virtual $t$-quark and a virtual $W^-$ boson and
then the $t$-quark reabsorbs the $W^-$ boson and goes over into an $s$-quark.
The photon can be emitted from any of the lines of the diagram. The
contribution from these diagrams involves two elements of the CKM matrix, the
very well-known element $V_{tb}$ and the reasonably well known element
$V_{ts}$. Thus, reliable predictions are possible. For a recent review
see$^{20)}$.

The rare decays are of interest for a variety of reasons. They demonstrate the
importance of pingwin diagrams. They test our understanding of QCD corrections,
which enhance the decay probability by more than a  factor of two. It is also
amusing to compare the simple pingwin diagram, which is used for rapid
presentations, with the complexity of the real calculation necessary to obtain
reliable numbers. In the real calculations one starts with an effective
hamiltonian, which is a linear combination of about ten operators

\begin{equation}
H^{eff} = \sum_i c_i O_i.
\end{equation}
Then at some high scale one calculates the coefficients $c_i$ as if a simple
perturbative calculations were valid. This is known as matching. Then one uses
the renormalization group equations to find the values of these coefficients at
a scale $\mu$ of the order of $m_b$. In order to do that, one needs the
anomalous dimensions, which form a $10\times10$ matrix. A typical final result
is$^{34)}$

\begin{equation}
10^4 BR(B \rightarrow X_s \gamma) = 2.8 \pm 0.8,
\end{equation}
to be compared with the corresponding experimental result$^{33)}$ --- $2.3 \pm
0.7$. The main source of uncertainty in the theoretical prediction is the
uncertainty about the final scale, which should be used. Following Ali and
Greub$^{35)}$ one uses

\begin{equation}
\frac{1}{2} m_b \leq \mu \leq 2 m_b
\end{equation}
and includes the corresponding spread of the results into the uncertainty of
the prediction. The scale can be obtained from a higher order calculation, but
this calculation is so difficult that it has yet not been completed, though it
is in progress. The corrections of order $0(m_b^{-2})$ are believed to be below
10 \% and, therefore, not very important at present. Thus, in a few years the
theoretical uncertainty is expected to decrease significantly. A precise
comparison of the predictions of the standard model with the data for rare
decays is of great interest for people working on extensions of the standard
model. Already at the present level of precision many possibilities of going
beyond the standard model have been excluded.

A few other rare processes should be mentioned. The branching ratio for the
exclusive process $B \rightarrow K^*\gamma$ has been measured$^{32)}$.
Theoretical calculations can reproduce the result, but they involve assumptions
about the structure of the $K^*$ and, consequently, are more model dependent
than the calculations for the inclusive process. The process $B \rightarrow
X_se^+e^-$ is of interest, because in it the $\gamma$ produced in the process
$b \rightarrow s\gamma$ has non-zero mass and consequently our understanding of
additional formfactors can be tested. The branching ratio for this process has
not yet been measured. Finally, measurements of the rare processes involving
the decay $b \rightarrow d\gamma$ could be used to get information about the
$CKM$ matrix element $V_{td}$.

\section{Nonleptonic decays}

For nonleptonic decays of particles containig heavy quarks, the standard
reference is still the paper of Bauer Stech and Wirbel$^{4)}$. According to
the model presented in this paper ($BSW$ model), for two body decays two basic
processes contribute. In class I decays, the heavy quark decays and emits a
virtual $W$-boson, which goes over into a meson. Then the quark produced from
the initial heavy quark recombines with the spectator(s) quark(s) and produces
another hadron. The probability amplitude for class I processes is multiplied
by a phenomenological constant $a_1 \approx 1$. In class II decays, after the
decay of the heavy quark the virtual $W$-boson decays into a quark-antiquark
pair. The antiquark from this pair forms a hadron by recombination with the
quark produced from the initial heavy quark and the quark forms another hadron
by joining the spectator(s). The probability amplitude for class II processes
is multiplied by a phenomenological constant $a_2$, which in absolute magnitude
is much smaller than $a_1$. Finally, class III decays can proceed via both
mechanisms. Their study makes it possible to determine the relative sign of the
constants $a_1$ and $a_2$. For a recent determination and discussion of these
constants see$^{36)}$.

The $BSW$ model becomes particularly simple, when the final state interaction
between the final hadrons is neglected. In this approximation, which is known
as the factorization approximation, the amplitudes for the nonleptonic two-body
decays can be expressed in terms of amplitudes for simpler processes. The decay
of the heavy quark is described by the amplitude occurring in the corresponding
semileptonic decay and the hadronization of the $W$-boson into meson $M$ is
described by the amplitude of the decay of meson $M$ into leptons. Thus, except
for the phenomenological constants $a_1,\;a_2$, all the dynamical information
necessary to describe the decay $H_i \rightarrow H_f M$ is contained in the
decay constant of meson $M$ and in the matrix elements $<H_f|J^\mu|H_i>$ for
the decay of hadron $H_i$ into a lepton pair and hadron $H_f$. This is
immediately seen from the graph for class I decays, but it can also be made
plausible for class II decays, which implies it for class III decays. Much work
is done on the problem, to what extent factorization is a good assumption. When
meson $M$ is fast, as is the case when it is a light meson e.g. a pion, it
leaves the interaction range before it has time to interact. This effect is
well known from discussions of the formation zone in scattering on nuclei. A
more formal discussion has been given by Dugan and Grinstein$^{37)}$. Data
have been analysed e.g. by Rieckert$^{38)}$ and this prediction has been
confirmed. On the other hand, a counterexample against the factorization
assumption is provided by the result$^{9)}$

\begin{equation}
BR(D^0 \rightarrow \overline{K}^0 \phi) = (0.83 \pm 0.12)\times10^{-3}.
\end{equation}
The quarks are, in the initial state $c,\;\overline{u}$ and in the final state
$s,\;\overline{d},\;s,\;\overline{s}$. According to the factorization
approximation to the $BSW$ model such a process is impossible, because at least
one of the initial quarks (the spectator) must be present also in the final
state, which here is not the case. On the other hand, taking into account the
final state interaction, one finds the two step process

\begin{equation}
D^0 \rightarrow \overline{K}^{*0}\eta \rightarrow \overline{K}^0 \phi.
\end{equation}
The first step can be described by the $BSW$ model with the factorization
aproximation, while the second is an allowed strong interaction process.
Thus, the problem is not so much to prove or disprove factorization, but to
find its validity range and practical methods to go beyond it.

Nonleptonic inclusive decays have been recently reviewed by Bigi$^{16)}$.
Let us mention two problems, which may indicate difficulties for the theory,
though in both cases certainly much work is needed before a firm conclusion can
be drawn.

There were worries that the semileptonic branching ratios for $B$ mesons come
out from theoretical calculations significantly larger than measured in
experiment$^{39)}$. This could be traced to an underestimate of the nonleptonic
decay widths in theoretical calculations. Careful evaluations of the decay
probabilities $b \rightarrow c\overline{c}X$ taking into account the finite
mass of the $c$ quarks and estimating the uncertainties of the
calculation$^{40),41),42)}$ have shown that one can easily reproduce the
experimental result at the expense, however, of predicting a larger decay width
for the decays $B \rightarrow c\overline{c}X$ than reported by
experimentalists. The discrepancy is below 3 standard deviations, but the
problem certainly deserves further study.

Another problem concerns the life time of the baryon $\Lambda_b$. Using a
suitable heavy quark expansion one finds$^{16)}$ that the spectator model
should be a very good approximation for the inclusive decays of hadrons
containing $b$-quarks. According to the spectator model, the life times of all
these hadrons should be equal to each other and equal to the life time of the
$b$-quark. An estimate of the corrections to this approximation has been
made$^{16)}$ and one finds in particular

\begin{equation}
\frac{\tau_{\Lambda_b}}{\tau_{B_d}} \approx 0.9,
\end{equation}
while the experimental value for this ratio$^{19)}$ is $0.72 \pm 0.06$. Thus,
once more the dicrepancy is of about three standard deviations.

As time goes on, one sees many such difficulties of the standard model
appearing and disappearing, but since most physicists hope for new physics to
make itself finally visible, it is interesting to know, which points now seem
to deserve particular attention.

\end{document}